# 9    Big Data and Geospatial Analysis
*Guy Lansley, Michael de Smith, Michael Goodchild and Paul Longley* (2018)

Perhaps one of the mostly hotly debated topics in recent years has been the question of "GIS and Big Data". Much of the discussion has been about the data: huge volumes of 2D and 3D spatial data and spatio-temporal data (4D) are now being collected and stored; so how they can be accessed? and how can we map and interpret massive datasets in an effective manner? Less attention has been paid to questions regarding the analysis of Big Data, although this has risen up the agenda in recent times. Examples include the use of density analysis to represent map request events, with Esri demonstrating that (given sufficient resources) they can process and analyze large numbers of data point events using kernel density techniques within a very short timeframe (under a minute); data filtering (to extract subsets of data that are of particular interest); and data mining (broader than simple filtering). For real-time data, sequential analysis has also been successfully applied; in this case the data are received as a stream and are used to build up a dynamic map or to cumulatively generate statistical values that may be mapped and/or used to trigger events or alarms. To this extent the analysis is similar to that conducted on smaller datasets, but with data and processing architectures that are specifically designed to cope with the data volumes involved and with a focus on data exploration as a key mechanism for discovery.

Miller and Goodchild (2014) have argued that considerable care is required when working with Big Data — significant issues arise from each of the "four Vs of Big Data": the sheer Volume of data; the Velocity of data arrival; the Variety of forms of data and their origins; and the Veracity of such data. As such, geospatial research has had to adapt to harness new forms of data to validly represent real-world phenomena. Some of the challenges posed by Big Data are new, whilst others are longstanding in geographic research and have been exacerbated by the recent data deluge.

In an article entitled "[Big Data: Are we making a big mistake](#)?" in the Financial Times, March 2014, Tim Harford addresses these issues and more, highlighting some of the less obvious issues posed by Big Data. Perhaps primary amongst these is the bias that is found in many such datasets. Such biases may be subtle and difficult to identify and impossible to manage. For example, almost all Internet-related Big Data is intrinsically biased in favor of those who have access to and utilize the Internet most. The same applies for specific services, such as Google, Twitter, Facebook, mobile phone networks, opt-in online surveys, opt-in emails — the examples are many and varied, but the problems are much the same as those familiar to statisticians for over a century. Big Data does not imply good data or unbiased data; moreover Big Data presents other problems — it is all too easy to focus on the data exploration and pattern discovery, identifying correlations that may well be spurious as a result of the sheer volume of data and the number of events and variables measured. With enough data and enough comparisons, statistically significant findings are inevitable, but that does not necessarily provide any real insight, understanding, or identification of causal relationships. Of course there are many important and interesting datasets where the collection and storage is far more systematic, less subject to bias, recording variables in a direct manner, with 'complete' and 'clean' records. Such data are stored and managed well and tend to be those collected by agencies who supplement the data with metadata and quality assurance information.

An important part of the geospatial analysis research agenda is to devise methods of triangulating conventional 'framework' data sources — such as censuses, topographic databases or national address lists — with Big Data sources in order that the source and operation of bias might be identified prior to analysis and, better still, accommodated if at all possible. Yet such procedures to identify the veracity of Big Data are often confounded by their 24/7 velocity. For instance, social media data are created throughout the day at workplaces, residences and leisure destinations, and so it is usually implausible to seek to reconcile them to the size and compositions of night time residential populations as recorded in censuses. As Harford concludes: "Big Data has arrived, but big insights have not. The challenge now is to solve new problems and gain new answers – without making the same old statistical mistakes on a grander scale than ever."

*"The recommended citation of this Chapter is: Lansley G, de Smith M J, Goodchild M F and Longley P A (2018) Big Data and Geospatial Analysis, in de Smith M J, Goodchild M F, and Longley P A (2018) Geospatial Analysis: A comprehensive guide to principles, techniques and software tools, 6th edition, The Winchelsea Press, Edinburgh"*

## 9.1 Big Data and Research

The academic community has been enticed by the potential to harness Big Data to predict real-world phenomena. No longer is research restricted by an inability to efficiently collect large quantities of data. In many settings Big Data offers comparable statistics to those traditionally collected by governments as officially released datasets, whilst in other settings they offer insights into phenomena that were not previously recorded. Efforts to gain access to new forms of data for research purposes are also motivated by an increasingly acknowledged necessity to extend the breadth of data sources available for research since traditional forms of data are costly to collect and in many cases suffer from decreasing response rates (Pearson, 2015). In many contexts, working with Big Data is a necessity. For population data, in particular, the traditional gold-standard sources of data are in decline around the world. Most developed countries are cutting back on their long-form censuses or replacing them altogether with administrative data (see Shearmur, 2010). In very important respects, Big Data can offer far greater spatial granularity, attribute detail and frequency of refresh. But these compelling advantages of greater content accrue without any guarantees that the coverage of the entire population of interest is known – in many Big Data applications reported in the literature, the population of interest may not even be defined! The analyst community may need to become increasingly reconciled to use of new data sources, and may certainly benefit from their richness; however, data analysts must also undertake increasing amounts of due diligence in order to establish the provenance of such sources.

There is a growing amount of interest in harnessing Big Data beyond their primary applications since they offer the potential to provide fresh insights into a wide range of phenomena across space and time (Dugmore, 2010). Furthermore, they offer large volumes of data and may have a good coverage of particular populations or spaces. In many settings, the insights that Big Data may be able to offer could not be generated by traditional data sources without being very costly. See, for example, Canzian and Musolesi's work on mental health using mobility data generated from mobile phones (2015) or Shelton *et al.*'s study of religion using the geoweb (2012). Big Data can present an opportunity to break away from a reliance on data of low temporal frequency to new forms of data that are generated continuously with precise temporal and spatial attributes.

Many datasets are generated in real-time by citizens using handheld technologies. Resultant data sources include mobile phone call records, WiFi usage and georeferenced social media posts. The velocity of their data generation allows very large volumes of geospatial information to be uploaded every day – this is a crucial step away from the slow production of geographic datasets and the delay between data collection and dissemination. Furthermore, the unrestricted spatial and temporal nature of data collection allows social research to be unshackled from an exclusive fixation on residential level data. For instance, Figure 9-1 shows the spatial distribution of Tweeters who submitted data between 8 am and 9 am on weekdays in 2013 in Central London. It can be observed that a large proportion of these users were using transport routes, many of which could have been commuting to work and therefore could be indicative of the spatiotemporal distribution of the population at large to a certain extent. Such information could not be acquired for large numbers of persons from traditional data sources.

Representation is a fundamental part of scientific knowledge discovery and is crucial for advancing our understanding of the world. All data should be thought of as a partial and selective representation of the totality of real-world phenomena, and the basis to selection should be understood. Data on places, people or activities are usually compressed into a set of digitally recorded variables in order to be stored efficiently. Therefore, representation is constrained by the scale and scope of each dataset, and the extent to which it depicts the totality of all real-world phenomena of interest. Most of what we know about the population and their activities has historically been dependent on generalizations and descriptions derived from very limited data. The skill of the scientist has been in devising sample designs that may found representations upon very sparse samples that nevertheless facilitate robust and defensible generalizations.

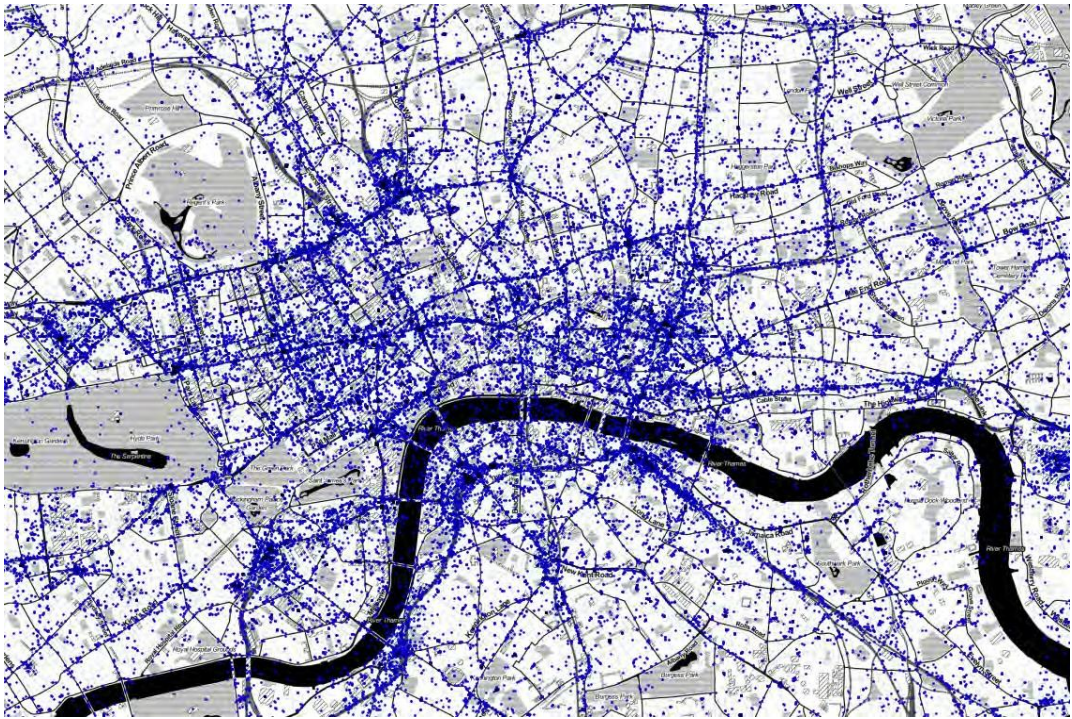

*Figure 9-1. The spatial distribution of Tweets in Central London sent between 08:00 and 09:00 on Tuesdays, Wednesdays and Thursdays in 2013*

In some fields of geography we are used to being data-rich. For instance, the Landsat program was producing exabytes of remotely sensed images in the early 1970s at such a velocity that the data were extremely challenging to handle. With improvements in technology, the supply of data on the physical environment has increased and improved in precision. However, data on people and their activities has historically been scarce. Now most big datasets are generated from human actions directly or indirectly and have countless applications for understanding the real-world. Yet, new forms of Big Data are fundamentally distinct from traditional datasets. Often their primary objectives are to make recordings of events and characteristics to assist administrative functions and there is little consideration of coverage as this is extremely costly. While on the one hand this means data are captured about what people actually do, rather than what they say they do in surveys, these actions will influence both who is represented in the data and how trends are manifested. Academics have little or no influence over the data collection process and many of the administrative processes of datasets remain hidden.

The diffusion of data collection into everyday activities has drastically enhanced the scope of data analysis. Firstly, in many cases, Big Data may represent a near-total coverage of specific populations, activities and places. Secondly, Big Data has enabled spatial analysis to shift large-scale analysis from aggregate data to individuals. For example, London's travel card data represents the vast majority of journeys that occur on public transport across the city, and each journey is recorded individually with origin, destination, and times appended to the recording. Moreover, some datasets also record the precise locations for individual records through GPS technologies. Individual-level data enables analysis to separate itself from issues associated with small area aggregations (e.g. the ecological fallacy and the modifiable areal unit problem discussed in section 4.2). Furthermore, often individual records can be linked to unique identifiers (such as accounts, addresses, etc.) enabling the conflation of large volumes of longitudinal data. It has also been possible to harness new technologies to track anonymized individuals across time and space in order to improve transport provision. For example, mobile phone app data can be used to estimate where and when a customer boards and departs buses in order to improve understandings of public transport use (Figure 9-2).

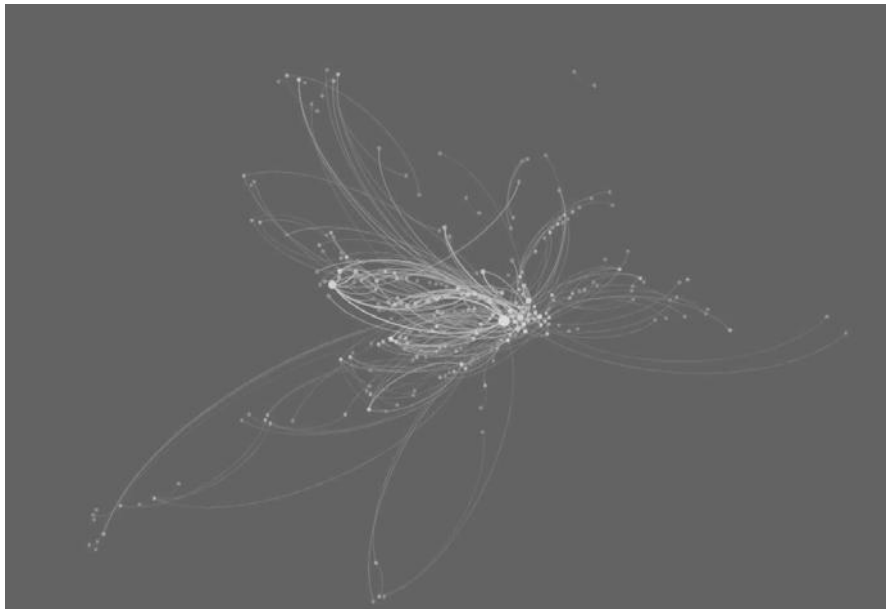

*Figure 9-2. A network graph to illustrate the bus stop to bus stop flows in Norwich, UK. Source: Stockdale et al, 2015*

These characteristics have led Big Data to change how we discover information from data. Previously, geographic research was grounded in theory partly due to the poor availability of data (by today's standards). Indeed many theories of spatial analysis are generalizations built on assumptions in order to estimate trends for situations where there are no or limited data. For example, spatial interaction models have been widely used to estimate movements in a broad range of applications. In the context of retail geography, their use is primarily to estimate the likelihood of persons from each spatial unit visiting each of their local retail destinations. Essentially, the choice to visit a particular destination is based on convenience (inverse distance) and attractiveness (sometimes simply the retail floor space). However, retail loyalty data can tell us the addresses of very large samples of patrons, as well as the times and dates when they visit stores. While one of the main criticisms of positivistic research is that models assume all individuals act rationally, through discrete data on populations researchers can also represent those that act irrationally in some senses. Indeed many actions will not conform to the generalizations that models are built from. Thus, large volumes of data have enabled us to mine trends and create predictive models with high levels of success.

As will be described later in this section, Big Data are not devoid of uncertainty especially when they are used to represent the real-world. Therefore, empiricist approaches may succumb to fallacies when the data collection procedures are not thoroughly understood. By letting data speak for themselves researchers neglect key influential geographic concepts pioneered by Geographic Information Science. Instead, data-driven approaches are limited to the identification and description of processes, rather than understanding why they exist. Theory is still needed to support critical research as Big Data remain only partially representative and research therefore may be founded upon data that are rich and detailed, but nevertheless not fit for purpose. It is important to think of most Big Data sources as a by-product, or 'exhaust' from a process that does not have re-use of data for research purposes at its heart. For example, a store loyalty card profile is a by-product of one or more transactions entered into by a consumer: it records the effectiveness of a company in delivering a service and how it makes a profit. It does not, however, record the entirety of that individual's consumption, and provides only indicators of a broader lifestyle that requires supplementation or validation from other sources for more broad-based research. As such, we must be cautious of over-reaching beyond the phenomena that the data may reasonably be purported to represent. This was demonstrated by the infamous over-predictions of Google Flu Trends, see for example Butler (2013).

## 9.2  Types of Big Data

In this section we discuss Big Data sourced from human interactions (e.g. social media platforms), process-mediated data, and finally machine generated data (e.g. from automated event and activity tracking).

### 9.2.1  Human-sourced data

Human-sourced information is integral to a wide range of services, some of which are explicitly for the dissemination of public information. For example, Internet enabled handheld devices have supported the growth of various Volunteered Geographic Information (VGI) services. A popular example is OpenStreetMap (OSM), which is a widely used crowd-sourced map of the world and geodata portal ([www.openstreetmap.org](www.openstreetmap.org)). Like other information sharing services (such as Wikipedia), the service uses a series of social hierarchies of moderators in order to maintain the quality of its information and sieve out redundant records (Goodchild, 2013). An advantage of VGI over officially produced counterparts is transparency and the velocity of data production. It is easy for lay users to update data and report misinformation, a process that is otherwise cumbersome with governments and businesses that often lay out lengthy internal administrative procedures. However, there are, of course, data quality issues that may emerge due to potential weaknesses in vetting the large quantities of data that are produced daily.

One very prevalent Big Data source is social media. Social media has become an integral component of modern societies. It is estimated that just under one-third of the world's population are currently social media users (Statista, 2017). There are several different types of social media including social network services, video sharing services, and information sharing services, most of which are accessed via online platforms. Consequently, social media has also become a popular source of data on the human condition for the academic community, largely due to the volume and velocity of the data and the rich social information they provide. Furthermore, many social network services such as Twitter and Flickr enable users to include coordinates from their mobile devices when uploading content, thereby enabling the study of geospatial phenomena.

As users engage with the online services through handheld devices, so they leave geographic and temporal footprints of their activity in the real-world within social media data (Blanford *et al.*, 2015). It is, therefore, possible that large quantities of social media data can be highly informative of general human activity and, thus, they have many potentially useful applications in geospatial analysis. Taking Twitter data as an example, many studies have sought to use the geosocial data to predict real-world trends, thus treating the users as sensors (Haklay, 2013). This has included estimating the spread of influenza (Lamb et al., 2013), predicting customer catchments for retail centers (Lloyd and Cheshire, 2017), and tracking natural hazards (Guan and Chen, 2014). The data can also be repurposed as indicators of urban activity by using new techniques in text mining to categorize Tweets and identify routine spatial and temporal patterns. For instance, Figure 9-3 displays the relative density of Tweets about education in inner London: here, most areas of high concentration coincide with the locations of university campuses.

Although social media data are curated by companies, they are distinct from most other commercial datasets in that the bulk of the data are fashioned by members of the public. The data typically record digitized media and can take many forms. Social network data are unlike other forms of Big Data as there is little control or regulation over what is produced, thus it is inherently difficult to harvest objective research (see Tinati et al., 2014).

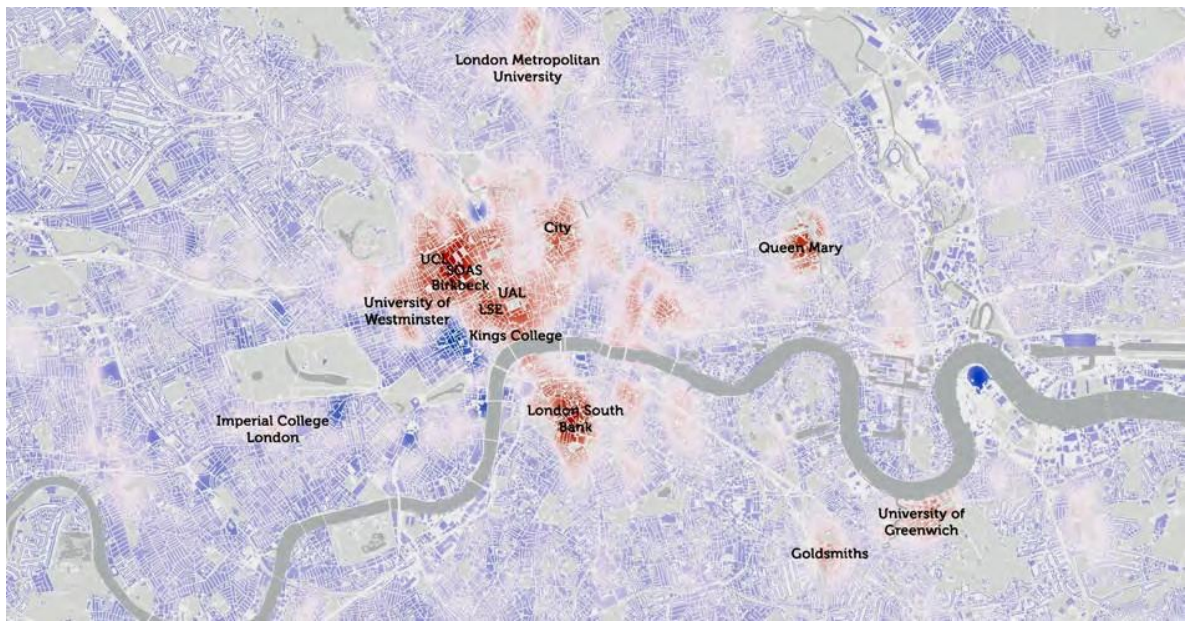

*Figure 9-3. The relative density of Tweets about education across inner London (20km East-West). Red indicates locations with higher densities. Source: Lansley and Longley, 2016*

## 9.2.2    Process-Mediated data

Process-mediated data describes the classes of data that are primarily generated through the transactions of businesses and governments. Such data is said to be "found" rather than "made", as they were typically generated to support commercial or administrative functions rather than to represent the real-world in research projects (Connelly *et al.*, 2016). Through new technologies, businesses and institutions now capture data at much higher velocities. Some example data sources are described below.

### Government Administrative Data

Administrative data are routinely collected by public institutions in the form of registrations, transactions and records. These include data generated by welfare, taxation, licensing and electoral registration, as well as records maintained by healthcare and education institutions. Most administrative data sources aim to acquire complete coverage of specific populations; and in some cases, the data are collected from transactions which are legal requirements, making them more suitable for geodemographic representations. For instance, governments usually retain records on 100% of vehicles and legally employed workers for taxation purposes.

Administrative data systems have historically been integral to the development of spatial data. It was for administrative purposes that the UK postcode and US ZIP code systems were built, and through them, Census geographies were carefully designed to assist dissemination at a fine level without breaching guidelines on disclosure. Due to common interests, much of the data collected by governments correspond with social surveys historically collected by statistics authorities. Consequently, on both sides of the Atlantic, governments are becoming increasingly interested in using administrative data to bolster census statistics (see Public Administration Committee, 2014 for example). Administrative data has the advantage of being routinely captured and regularly updated, enabling large-scale longitudinal analysis of high quality.

In addition, the data can contribute new information that were not previously available in the public domain. For example, despite considerable interest in social class and wealth from social scientists and users of geodemographics classifications, there are still no granular data on income in the UK. Previous research has found value in harnessing government data to investigate life-chances and criminality (Britton *et al.*, 2015). In addition, while previous geographic research on deprivation has included car ownership as a proxy for social standing due to its collection in censuses (for example the Townsend Index of Deprivation, Townsend *et al.*, 1988), research on car registration data from the UK's Diver and Vehicle Licensing Agency found car characteristics (such as age and model) to be strongly associated with socio-economics, especially in large cities. Figure 9-4 demonstrates that there is a distinctive geography to the registered locations of new cars.

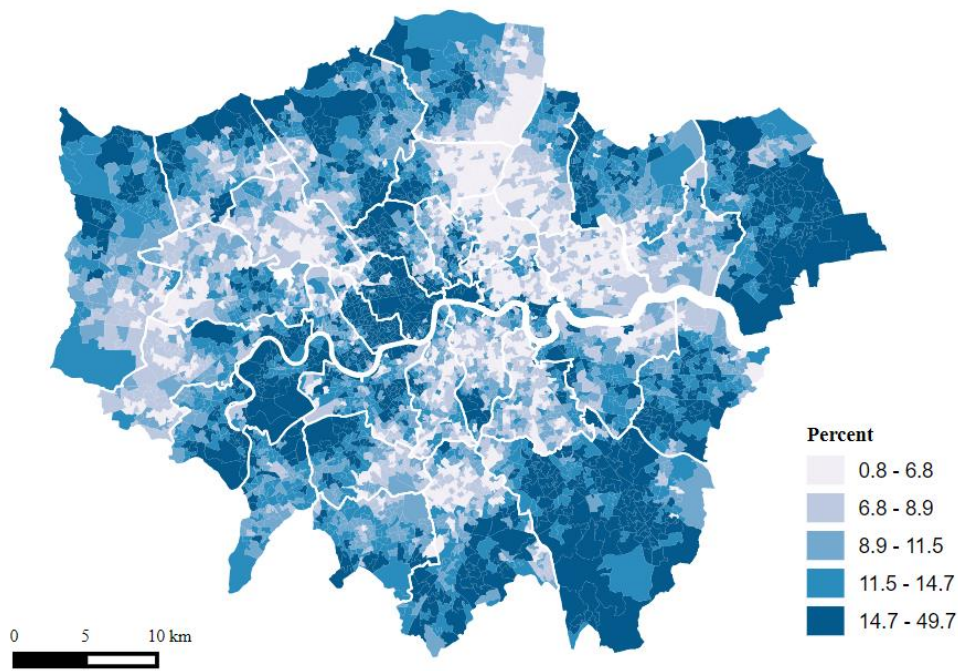

*Figure 9-4. The spatial distribution of the ratio of vehicles that are less than 3 years old by their registered addresses. Source: Lansley, 2016*

Of course, the extent and quality of administrative data collection varies around the world. Some countries maintain detailed population registers in order to maintain better data on citizens. Such registers enable different government institutions to link their data in order to improve service delivery. Unfortunately, in most countries, administrative data are collected by multiple different organizations making it difficult to link records, although there have been efforts to overcome this using multivariate linkage approaches.

## Consumer Data

An increasing share of data held on people and their actions is generated by commercial organizations in order to assist their activities. These include data created by retailers, utility providers, transport providers and banking and financial services. Typically, they require transactions of some kind to generate data although they often also retain account information. For example, retailers record the times and the locations of transactions. Some large retailers and almost all online retailers also maintain customer accounts. Sometimes, specially devised loyalty programs (such as supermarket loyalty cards) have been implemented specifically to understand trends in consumer behavior and to target promotions. This enables retailers to retain valuable longitudinal data often including the location of shopping trips and customer residences. Loyalty databases and website shopping accounts empower retailers to focus their data collection and promotions in order to target individual customers.

A popular anecdote about the predictive capabilities of retail data describes how Target (an American discount store) predicted a teenage customer in their loyalty database was pregnant (Duhigg, 2012). Target used their databases to personalize the coupons and advertisements they mail to their customers. In this case, a father complained to the store that his daughter had been sent promotions for baby products and maternity clothing. His concern was that the retailer was implementing marketing techniques that might encourage teenagers to have children. However, a few days later the father contacted Target again to apologize — he had discovered that his daughter had been pregnant for some time. It is not uncommon for private companies to use empiricist methodologies in order to automate tailored service provision at the individual level. In this case, the high dimensionality and granularity of the data made it feasible to make associations at the individual level.

Some large retailers may achieve a very large coverage of the population. For example, Tesco's Clubcard in the UK had achieved a coverage of 15 million members in 2015 (over 30% of the adult population). However, engagement with the program will vary considerably as many users may become disengaged with using loyalty schemes over time or may routinely avoid using cards when making small purchases. Yet, it is obvious, that this data will have merits over traditional data sources that will struggle to acquire detailed information on consumption for large shares of the population. Retailers are now able to grasp data on the consumer behavior of groups who are traditionally more averse to participating in market research program (Manyika *et al.*, 2011).

Retail data, therefore, can be of considerable value to geospatial population research. The core use of geodemographic classifications in industry is to predict lifestyles across space, particularly consumer behavior (Harris *et al.*, 2005). Consumption habits may have broader connotations for understanding segregation in lifestyle choices. Indeed, equivalent data from traditional sources suffer from common limitations of costs, respondent errors, sample biases, and perhaps most importantly, small sample sizes.

### 9.2.3   Machine-Generated data

Machines and sensors are routinely used to capture events, measure consumption or record the physical world. The Internet, and technological innovation more generally, has enabled sensors to capture a wide variety of data in a non-disruptive way. Examples include fixed sensors which are in place for security or monitoring purposes, mobile sensors which use GPS technology to capture spatial and temporal trends, and also computer systems. One of the reasons that we are able to gather such vast quantities of data at a high velocity is that we have decreased the role of humans in data generation. It is now easy to generate accurate and precise spatial information from mobile devices — previously georeferencing was expensive and far more complicated to implement. Together, machines generate several different types of new forms of data that can be used to assist decision making in real-time. For instance, "Smart Cities" amalgamate real-time data from a wide range of sensors including CCTV, weather monitors and traffic measuring sensors, in addition to other types of Big Data (such as social network data).

An example of fixed sensors are machines which are used to monitor energy usage. Recently, UK energy companies have begun making the move toward installing smart meters for every property. These devices record energy consumption in real-time enabling the collection of energy trends for addresses at a very high temporal frequency. Most commonly, data are stored in 30 minute intervals. While the primary objective of the data is to improve the efficiency and transparency of billing, they can also reveal interesting geographic trends. For instance, Dugmore (2010) suggests that as occupied households use utilities daily, unoccupied dwellings could be identified by pooling utilities companies' data. The high temporal resolution of the data enables us to segment households based on energy usage throughout the day. Research has found smart meter data to be an effective predictor of household characteristics and a viable alternative to the Census for some measures. For instance, Samson (2014) clustered the standardized average daily energy profiles of smart meters from deprived postcodes. The research identified that most energy profiles fit one of four trends (Figure 9-5). Each of the trends were found to be associated with households of different life stages.

There are also sensors that are unfettered from physical locations. Satellite imagery has historically been a common source of geographic data and there has often been more data than researchers can handle. However, there are now new forms of data that have benefited from GPS technologies. One such example is fitness apps that use mobile devices to record routes taken during exercises (Figure 9-6). These types of apps (and indeed many others) record detailed tracks of sensor movements and can be informative of real-world travel activities. Typically the apps record locations at very regular intervals so that journeys can be mapped and attributes such as elevation and speed can be appended. The proliferation of handheld devices has vastly increased the supply of geospatial data on the population. It is not uncommon for such data to be re-purposed to estimate traffic congestion and the popularity of restaurants and stores.

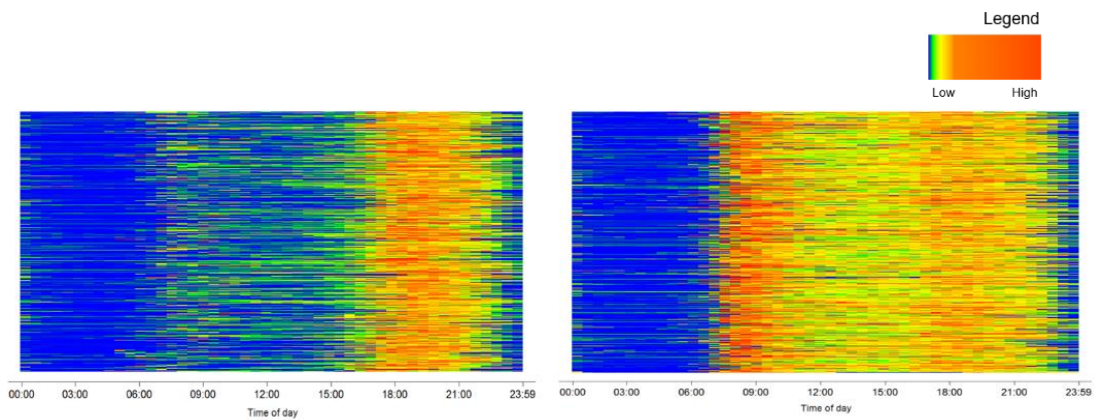

*Figure 9-5. Temporal heat maps of energy usage across a typical day by four clustered energy profile types. Each row is an anonymized smart meter for a unique household. Source: Samson, 2014.*
*(Continues on next page)*

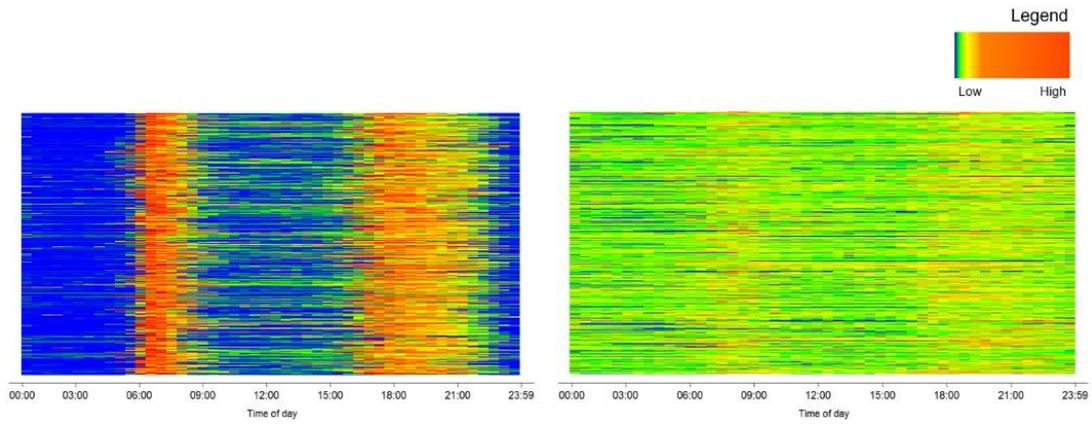

*Figure 9-5. Continued*

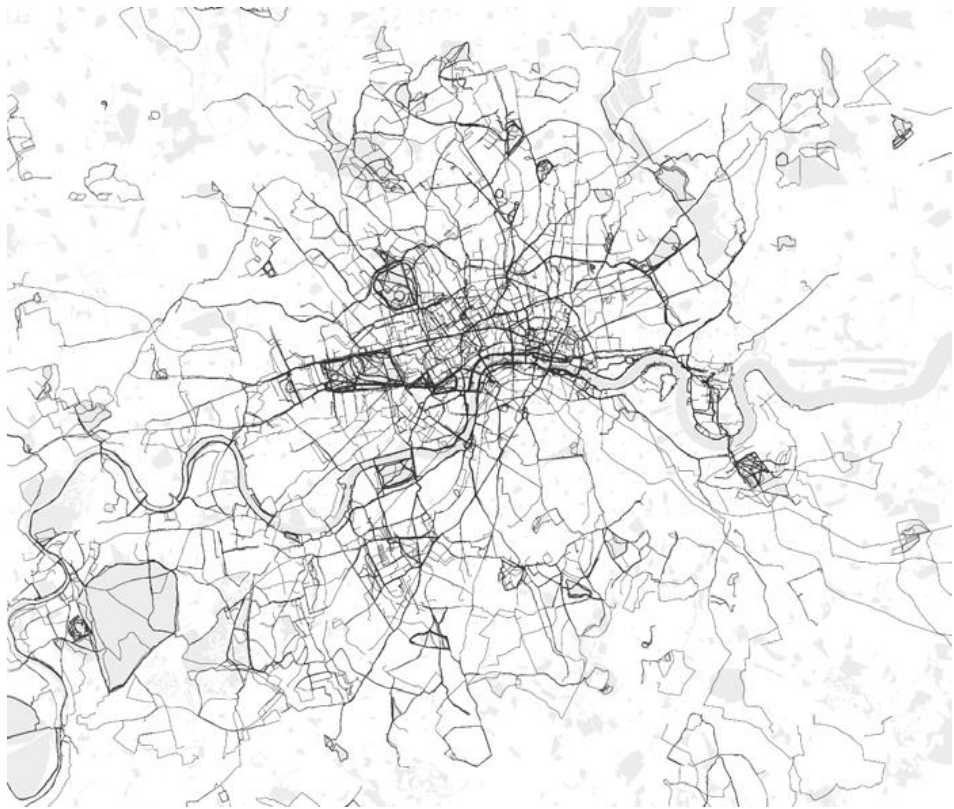

*Figure 9-6. Running routes recorded by a fitness app in London (30km East West). Source: Lansley and Cheshire, 2018*

## 9.3 Challenges of Big Data

Unfortunately, despite the enthusiasm for new forms of Big Data, there has been relatively limited critical research into their validity for geospatial research on real-world phenomena. This is foremost because the provenance of big datasets is often unknown and it may be difficult to link records to alternative databases. In addition, the unstructured and unregulated nature of some datasets has deterred some from using them at all. Whilst Big Data are useful for countless applications, it is important to be cognizant of their limitations. This section describes the core fundamental challenges to harnessing Big Data to form geospatial representations. It is important to consider that these challenges may not apply to each data source to the same extent. Each big dataset will have its own merits and deficiencies that researchers should seek to explore.

### 9.3.1 Access

Data are unlike other assets in that they are difficult to collect but easy to copy and disseminate. However, only a very small proportion of big datasets are released in the public domain due to their strategic importance, commercial value and potentially disclosive attributes. Breaches of privacy can have serious adverse consequences on the public opinion of businesses and governments, so there are considerable efforts to safeguard data. The recent events surrounding the misuse of Facebook profile data by an academic and Cambridge Analytica only serve to highlight how fraught with problems issues of access and privacy can be. This inevitably has had a detrimental impact on Big Data research as commercial organizations that produce geospatial Big Data products often value efficiency and customer satisfaction over other interests. Companies are also less likely to share their methodologies and results thus hampering the advancement of research, methodologies and theoretical insights.

Where data access is granted, researchers and analysts are usually at the mercy of their data providers. It is not uncommon for elements of the data collection or sampling process to be kept private due to commercial sensitivities. Many companies that opt to share samples of their data disseminate them via Application Programming Interfaces (API). However, when using such channels the coverage of the data are usually uncertain (Boyd and Crawford, 2012). Taking the example of Twitter, little is known about how the free data feeds are sampled and feeds often drop in velocity for short intervals of time. Furthermore, data owners may set conditions on data access and use, and sometimes these limitations may be contrary to the interests of researchers.

For these reasons there needs to be greater support at all levels for data sharing and safe data practices in order to facilitate research and maximize the full potential of Big Data. In 2008, the OECD launched a call in support of open data aimed at national governments. In response, some governments have launched open data initiatives to disseminate administrative data which may be of value to research and the general public (for instance data.gov in the USA, data.gov.uk in the UK and aurin.org.au in Australia). Example datasets that have recently been made open source in the UK include Land Registry Price Paid data (the location and value of almost every domestic property sale) and Domestic Energy Performance Certificates (the energy performance certificates for over 15 million properties).

Commercial organizations are generally under less pressures to share their data. Therefore, without legislative intervention, institutions are required to provide safe environments through which commercial data can be accessed by researchers without violating data protection protocols. One such institution, is the UK Consumer Data Research Centre (CDRC, based at University College London) which specializes in large consumer datasets and manages data from a wide range of sectors.

## 9.3.2  Ethics

Whilst concerns about the use of Big Data have been raised by many parties, ethics are frequently overlooked by practitioners who have been enticed by new and innovative forms of insight (Puschmann and Burgess, 2014). The majority of Big Data production and analysis currently occur within the private sector. Given that data and algorithms are commercial assets, their methodologies are often trade secrets and thus not subject to independent scrutiny and critique. Data on people are extremely sensitive, both to the people they relate to and to the organizations that own the data. In 1995 Ground Truth was published to bring attention to the social implications of geographic information systems and geographic databases of people (Pickles, 1995). All of the chapters in this seminal collection express concern, to differing degrees, over the impacts of GIS, some of which focused on the negative connotations geographic research can have on society. However, the ethical concerns have in fact been exacerbated in recent years owing to the proliferation of data which are now analyzed at the individual level.

The collection of personal data is integral to the personalization of services (as demonstrated by the Target loyalty program described above). However, the scope and use of Big Data can result in "dataveillance", a process through which individuals' details are routinely collected and stored creating detailed databases about people and their behavior (Dodge and Kitchin, 2007). It is therefore often possible to identify individuals and their behaviors from pooling different sources of Big Data. In addition, some data present very detailed footprints of user locations making it possible for analysts to distinguish personal characteristics about individuals (including their place of residence; Valli and Hannay, 2010). Therefore, when data outputs are publicly disseminated, efforts are undertaken to ensure that data are not disclosive of individuals. Usually, this entails aggregating the data into predetermined spatial units. However, this means that in many cases, researchers cannot take full advantage of the fine granularity of Big Data and must succumb to the challenges of ecological fallacy. It is therefore challenging to secure privacy in some datasets whilst retaining their quality and precision.

Data protection is becoming a more contentious issue and this is also being reflected by new legislation around the world such as the General Data Protection Regulation (GDPR) which is scheduled to come into effect in the EU on 25 May 2018. Once access has been granted, the researcher needs to ensure they are compliant with data protection procedures and also maintain a good understanding of the potential implications of working with big datasets. A useful resource is the British Academy and Royal Society's 2017 report entitled "[Data management and use: Governance in the 21st Century](#)" (Royal Society, 2017), which summarizes the ethical issues that arise from data in today's society and how the issues might be mitigated.

## 9.3.3  Data Quality

Data quality issues are paramount to big datasets and are one of the major criticisms that have deterred many academics. Errors can arise from all parts of the data generation and amalgamation processes, from measurement error to adjustment errors (see the total error paradigm in Groves *et al.*, 2011). These errors can then contaminate other data when records are linked and are inherently difficult to identify. Furthermore, the volume and velocity of data prevent any efficient means of validating records.

Small sets of data are typically created using procedures that are transparent and robust. However, most forms of Big Data are not subject to standard quality controls and they come from numerous different sources of varying quality and reliability. That said, most datasets never enter the public realm so there is little knowledge about how rigorous the collection and data cleaning procedures are. Within the commercial sector companies may focus on quantity over quality and emphasis is placed on efficiency in order to maximize profits. It is, therefore, unsurprising that Big Data may contain an unknown proportion of noise generated by errors and inaccuracies.

While most forms of Big Data are generated by machines that have reduced the elements of human error (especially with regards to temporal and georeferencing components), there are cases where machines can be inaccurate. For example, GPS coordinates are known to be less accurate in so-called urban canyons. Yet, as seen above, there are is still a large share of Big Data that are human-sourced. It is not uncommon for errors (accidentally or deliberately) to occur in VGI. For example, the quality of OSM data has been found to vary due to the skills and interests of contributors (Haklay, 2010). Furthermore, business and administrative data could also include errors that arise due to mistakes during processing. For example, the falsification of administrative records could lead to changes in policy and allocations of funding (Connelly et al, 2016).

Data quality also often deteriorates over-time as places and events are not static. For example, administrative and consumer databases regularly retain the incorrect residential addresses of adults for individuals who have moved. It is, therefore, important that each record is timestamped to improve the transparency of databases. It is also useful to synthesize data from multiple sources to reinforce confidence in observations and measurements and to identify outliers.

### 9.3.4 Repurposing Data

Often Big Data are used in research despite the fact that they may have been collected for completely separate reasons – this is particularly challenging when attempting to predict human actions. For example, research has found social media data to be of variable utility for predicting election outcomes. Whilst pollsters can ask the all-important question about voting intentions, researchers attempting to glean political predictions from social media must acquire a plethora of messages and then attempt to contextualize them. Big Data fundamentally represent phenomena on the population and their behaviors that is distinct from traditional social science datasets. There are advantages in this, for instance surveys are prone to misreporting (deliberately or accidentally) whereas in many instances Big Data are accurate due to the automation of data collection through machines. However, this is not to say that all data must, therefore, be valid and free of constraints. Big Data are not neutral and may be prone to misrepresentations in the context of geospatial research.

Whilst it is reasonable to assume that Big Data are reliable proxies for real-world activities or conditions, in making these assumptions it is important that we take account of the possible implications that the means of data collection may have on representations. Activities which generate data may be constrained by time and geography, they may also be unevenly distributed across the population (as shall be discussed later in this section). In addition, people may engage in the activities differently and it is not possible to obtain information on their motivations. For instance, social media contains an unregulated and unknown proportion of individuals' beliefs, opinions and activities. In addition, some data are generated by automated programs (bots). Moreover, the proportion of persons that engage with social media, and their level of engagement, is also unknown. It is, therefore, a challenge to link trends on social media to events occurring in the real world.

When generating inferences about the real-world from Big Data, it is possible to generate misrepresentative indicators. For instance, mobile phone data can leave traces of users' mobility at a relatively high spatial and temporal coverage. However, inferences of travel and user behavior from these footprints beyond their presence could be difficult. For instance, it may be reasonable to assume that is possible to make estimations on how crowds move from A to B, but for how many journeys would such models be accurate? Unfortunately, it is not often possible to conclusively validate models built from Big Data beyond links to aggregate datasets, small sample surveys or alternative big datasets with similar limitations.

### 9.3.5 Demographic Bias

Sampling has always been a fundamental component of geographic research. In order to devise nomothetic (law-based) knowledge, hypotheses need to be built from representative samples. Bias typically arises due to one of three reasons: the selection procedure is not random; the sampling frame does not account for all groups in society; and/or some groups are impossible to reach (Moser and Kalton, 1985). In the case of Big Data, they are usually a by-product of actions and therefore are not random nor scientifically sampled at all, and some groups may be excluded. Instead of building representations from carefully constructed sampling frames, researchers working with Big Data acquire data and then must attempt to work out who they truly represent.

One very early failure of a big dataset was when the Literary Digest's postal poll regarding the presidential election received roughly 2.4 million returns in 1936. With the aim of achieving as large a sample as possible, the magazine sought datasets which contained the names and addresses of millions of adults, these primarily comprised vehicle registration lists and telephone directories, and in total over 10 million ballots were posted. However, despite receiving an impressive number of responses, the poll incorrectly predicted that Landon would beat Roosevelt. Their data sources are now understood to have produced biased samples that were likely to be of a higher socio-economic status. The rates of both automobile and telephone ownership were much lower amongst poorer adults.

Generally, in both academia and the commercial sector, large sample sizes are valued as more favorable where they represent a greater proportion of the "population" of interest. Indeed, Poisson's law of large numbers entails that with larger samples one is more likely to acquire an observed mean closer to the theoretical mean of a given phenomenon. Despite the promises of Big Data that they represent complete populations ("N = All"), the majority of datasets unfortunately fail to capture every citizen. Data may represent all users of a given service, but does everyone in the population use that service? It is not unusual for a minority of the population to produce the majority of the data (Crampton *et al.*, 2013). Big Data are devoid of robust sampling frames meticulously designed to reduce sample errors and sample biases. In addition, the primary objectives of most big datasets are not to acquire complete coverage of the population at large, thus data are prone to representing particular subsets of the population that engage with the various activities that generate data. Moreover, it is very difficult to filter out incorrect information, often government and commercial bodies rely on the individual correcting their records themselves.

Whilst most representations are necessarily partial, incompleteness has previously been considered to be both systematic and known. A disadvantage of Big Data is that the data are often of unknown provenance. It is not uncommon for big datasets to contain no demographic variables at all making it inherently difficult to estimate their representativeness of the population at large. This is usually because asking participants to volunteer additional information may discourage them from using the particular service. We are, therefore, still reliant on censuses for good quality primary geodemographic data available at fine geographic scale, despite the fact they are collected very infrequently.

Data linkage is often required to ascribe demographic characteristics to Big Data, bolstering their utility and reducing bias. Linkage techniques in geodemographics generally fall into one of three categories: a) linkage to individual-level data, b) linkage to small area population statistics or c) inferences based on other personal variables in the data. Individual data linkage is the only means of acquiring accurate demographic variables without succumbing to issues of ecological fallacy. Unfortunately, individual-level demographic data are very rarely available to researchers due to their confidentiality restrictions. However, such linkage has been made feasible in some countries where individual-level data are routinely collected at vast scales, for example, medical research in Sweden.

Another means of estimating the demographic bias of geographic datasets is through data linkage to small area official statistics. However, in these instances, validity is limited by the quality of the official population statistics and is hampered by issues associated with spatial aggregations (for example, the scale and zoning effects (Openshaw and Taylor, 1979) and the ecological fallacy (Openshaw, 1984)). While some have taken advantage of microsimulation procedures in order to retain the focus of analysis at the individual level, they are still haphazard estimations based on assumptions.

An additional, but often overlooked means of linking demographic characteristics to big data is through linkage to datasets that record the statistics of personal attributes which are often recorded in large databases. A prominent example of such attributes is names. Names appear in many different big datasets that require users to register accounts (such as social media or store loyalty programs). Both surnames and forenames are likely to be distributed unevenly across a wide range of geodemographic domains. For example, most European names are sex-specific. Consequently, there is value in devising demographic databases on forenames in order to provide typical statistics on age and gender that can be ascribed to other data. Figure 9-7 demonstrates this by comparing UK Census data to estimated demographics from the names detected in a 2011 population register that has near complete coverage of the adult population. Although, of course, not every individual will share the modal characteristics of all bearers of the same name.

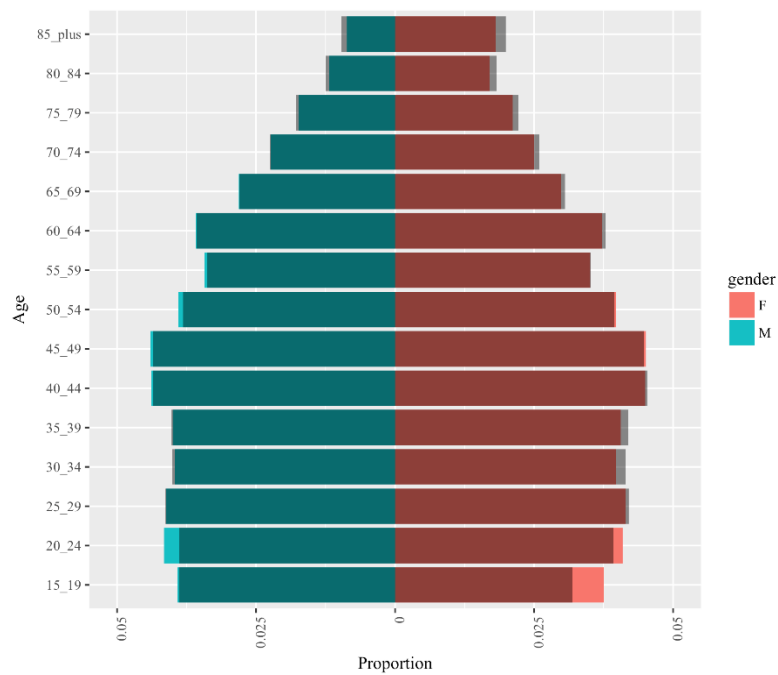

*Figure 9-7. Population pyramid based calculated from inferring demographics from the full names of residents recorded in the 2011 UK Consumer Register (colored) versus the equivalent demographics from the 2011 Census of Population (grey). Source: Leak, 2018*

## 9.3.6    Spatial and temporal Coverage

In addition to representing biased subsets of the population, data are usually partial representations of phenomena across space and time. Activity of some kind is usually required to generate data, and these are bounded by geography and time. Even when machines are used to automatically collect data, their geographic placement may not be uniform and may be influenced by the demographic biases described above. Most commonly the responsibility of recording data is handed to the people the data pertain to. Therefore, their behavior will heavily influence the data they generate and they may also be influenced by unobserved external factors. Failure to account for the geotemporal biases may lead to the misrepresentation of real-world processes.

Some new forms of data can be generated anywhere and at any time due to the use of new technologies such as hand-held devices. However, despite the presumed freedom of these data collection activities, they will still have some spatial and temporal restrictions. People cannot be assumed to actively contribute data randomly or at regular intervals – unless data are machine generated of course. Taking georeferenced social media data as an example, the action of uploading a georeferenced post may be more common in tourist or social settings than in workplaces. And indeed, some factors such as time of day/night or poor data communications will inhibit the ability to upload content. Therefore, distribution of georeferenced social media does not represent the distribution of people (and activities) across time and space, despite there being strong associations in some settings.

At the international level, Taylor and Overton's first law of geographical information still stands true today at a general level, viz: poorer nations are also poorer in good quality data (Taylor and Overton, 1991). There is a digital divide in a range of different types of geographical data. For example, the quality and precision of OSM is far better in developed parts of the world where there are more contributors and there is generally better connectivity. In addition, social media usage is generally lower in developing nations. Although Internet connectivity is improving globally, developing nations and those with lower information equality still lag behind. This is unfortunate, as the developing parts of the world are often in the greatest need of good quality geographic information.

### 9.3.7    Unstructured Data

Geographers have long attempted to make sense of intangible concepts. Non-quantifiable concepts are well suited to ethnographic researchers who are able to collect very detailed observations and derive conclusions based on shared experiences. However, it is a more novel challenge to represent intangible concepts from large quantities of data in research, as quantitative data typically represent phenomena as categorical, ordinal or interval, so that they can be analyzed numerically and efficiently.

One of the challenges with many types of Big Data is that often their records are not numerically structured making it inherently difficult to undertake quantitative analysis. Techniques are, therefore, required to manipulate the data. Taking the example of social media posts that are textual, often research has focused on isolating a series of keywords and then using filtered searches to generate relative frequencies across time and space. This has enabled researchers to use social media posts to track real-world events such as flu epidemics (Lampos et al., 2010) or earthquakes (Sakaki et al., 2010). In these approaches, each case is a binary datum; either a relevant term is present or it is absent (1 or 0).

Another set of techniques attempt to group words or entire social media messages in order to produce categories. By modeling the co-occurrence of particular words across large samples of documents, it is possible to create associations between words, and also, associations between posts. Topic modeling techniques have provided a popular means of analyzing large documents, such as blog posts, speeches or articles; however, their utility on social media is questionable due to the short length of documents and unstandardized use of language. Despite this, it is feasible to devise topic classifications from short social media posts if appropriate data cleaning steps are applied first. For instance, Figure 9-8 displays the relative frequency of geotagged Tweets from Inner London that were allocated to 20 key topic groups by hour of day. Some of the temporal patterns could be indicative of activities, for example the Food and Drink group shows peaks at mid-day (lunch time) and in the evening, as would be expected.

By converting the data into more reductive forms, there is an inherent loss of information. It is likely that qualitative analysis on a small sample of data may reveal entirely different trends and perspectives. In addition, quantitative analysis of large amounts of data tends to ignore the contextual elements of each datum. While it may be possible to group messages based on textual similarity, it is not possible to return meaning in a numerical form. Users describing the same topic may be tackling it from entirely different viewpoints with very different agendas. In the case of tracking flu from Twitter messages, it is easy to confuse messages reporting symptoms with those that are merely expressing awareness of the illness.

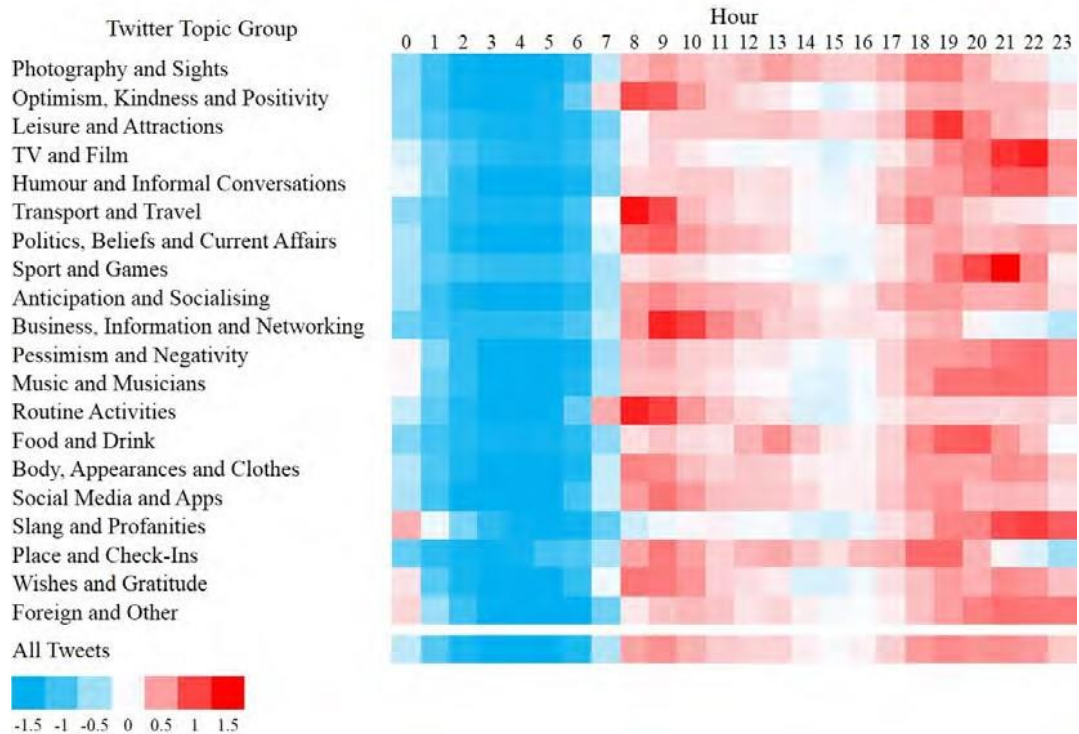

*Figure 9-8. A heat map of the temporal frequency of Tweet topic groups across the whole weekday sample by hour of the day. The data from each of the Twitter groups has been standardized as z-scores. Larger numbers (red) therefore indicate over-representation. Source: Lansley and Longley, 2016*

## 9.3.8  Data Linkage

Although rich in detail, most big datasets still only offer partial representations and typically only pertain to a very limited number of variables. Therefore, as discussed earlier, data linkage is required to improve the coverage of data and enhance their predictive capabilities (Harris *et al.*, 2005). Combining Big Data maximizes their value and enables critical research to ask questions about distinctive phenomena that may be extrinsically associated with each other. Having discovered the value of linked data, some countries have introduced data infrastructures which enable the linkage of individual records to facilitate research and public service delivery. However, it is not feasible to link most data from commercial organizations given that few efforts have been made to standardize how they collect information on individuals.

Yet even where consumer data are more widely available, challenges to effective concatenation and conflation remain, not least because of the ways that different data are structured by the organizations that collect them. In short, data linkage is plagued by uncertainty and the potential for propagating errors. There have been countless efforts to link big datasets, often using probabilistic linkage techniques. Such techniques cannot match every case and may often incorrectly match records (Goerge and Lee, 2001). However, without individual-level linkage data either remain isolated or trends are observed through aggregations and associations with small data.

Often unique identifiers are particular to each dataset and so linkage is reliant on other unique elements such as addresses and names. However, both of these can be input and structured inconsistently. Whilst postcode and zip code systems were designed to assist the identification of addresses, in the UK, for example, a single postcode covers 15 properties on average, often considerably more in urban areas. In addition, postcodes may change or be recorded incorrectly. Research has previously attempted to build a longitudinal database of the UK adult population from 20 separate annual population registers. It was not possible to link roughly half of the addresses using a simple string match due to inconsistencies in formatting and spellings. Therefore, a bespoke matching algorithm had to be devised. The same research also found that person names could be recorded inconsistently, thus hampering linkage. However, using specially devised heuristics, over 100,000 women each year were identified to have probably changed name following a marriage for instance. Indeed, with an understanding of how identifiers could be recorded differently, it is possible to tailor linkage algorithms to maximize match rates. Indeed, it was even able to estimate the origin and destination of over 750,000 domestic migrants by looking at subsets of households that joined and left addresses with unique combinations of names (Figure 9-9).

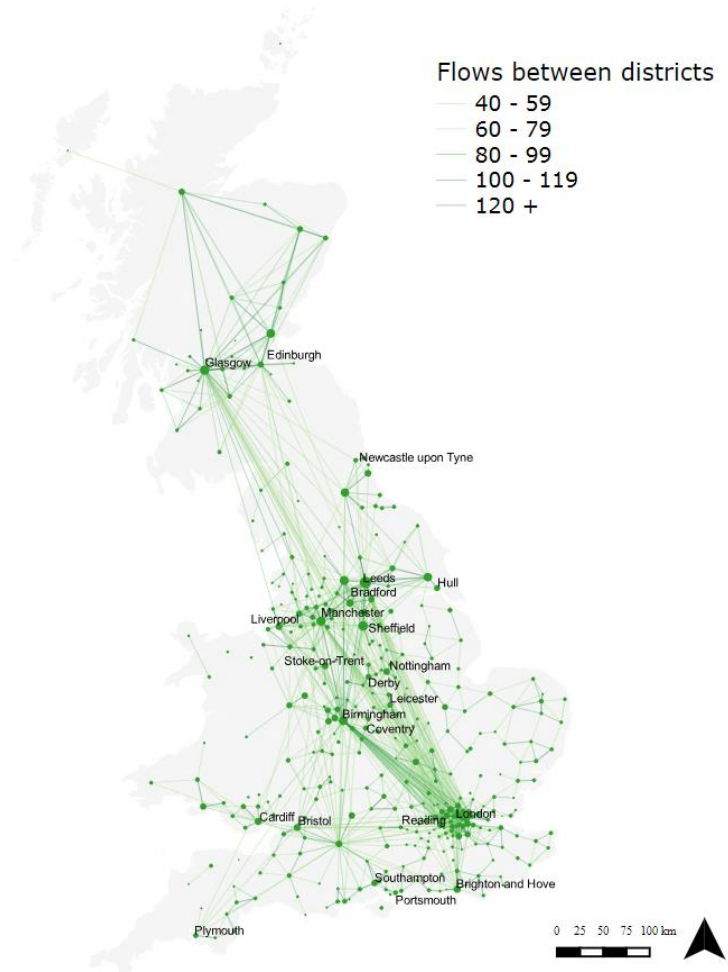

*Figure 9-9. The estimated migration flows between local authorities in Great Britain derived from novel data linkage techniques applied on the 2013 and 2014 Consumer Registers. Only flows of more than 40 people are shown. Source: Lansley and Li, 2018*

### 9.3.9    Tools and Skills

The volume, velocity and variety of Big Data have imposed considerable challenges on geospatial data analytics. Whilst there have been advancements in computer memory and databases, the majority of conventional analytics packages struggle to handle very large datasets. This is especially true for geographic information systems that also often need to account for complex issues such as spatial dependence and spatial non-stationarity (Shekhar *et al.*, 2011). Here more spatial data can equate to exponential increases in computing required to undertake spatial operations. The complexity and the extensiveness of some spatial operations (and in deed various multidimensional modeling techniques) means that it is not always feasible to employ parallel computing to increase their efficiencies. Indeed, most spatial statistics techniques were originally devised to work on a finite number of points.

Visualization remains an imperative component of quantitative analysis. Carefully selected techniques are required to convey trends at the expense of noise in Big Data. Too much detail can make key patterns unobservable. In addition, scale is still important, zooming in too far can obscure local trends. Therefore, despite the detail and precision of Big Data, visual outputs must remain reductive in order to support decision making. Whilst developments in static image-based representations have been minimal, there has been a growth in the implementation of web-enabled services (such as "[slippy map](#)" platforms) that let users explore the data and focus on key areas of interest.

Data on people and places are also generated faster than they can be critically explored and understood. It is also the case that most techniques in spatial analysis were built to work on static and pre-existing databases which are therefore treated as static and timeless entities (Batty, 2017). However, a large share of Big Data are generated at high velocities and there are interests in analyzing them in real-time (for instance the data generated by smart cities). Furthermore, the handling of such data is challenging, for instance indexing techniques that were designed to make very large datasets accessible can become inefficient once data goes over their original capacity of extension. In response to this, bespoke software packages have been developed to harness real-time geosocial information in order to detect real-world trends. There are also new spatial data streaming algorithms which were primarily used to estimate weather from new flows of Big Data. However, often data are statistics that are generated for data in intervals, thus the formation of intervals become a fundamental part of the analysis.

Data may also come in a variety of different types, and they may require extensive reformatting procedures in order to work with them. Geospatial data generally falls into three forms: raster, vector and graph, and there are established practices for working with all three. However, as observed in the Unstructured Data section above, many Big Data sources generate data types that are not conventional for quantitative analysis. While this has not changed how spatial elements of data are standardized, the appended data now vary considerably. These include textual data, images and even video feeds. Therefore, new techniques primarily from the field of computer science have been devised to harvest information from new forms of data.

In response to the technical challenges of working with Big Data, researchers have to become more comfortable working with coding and scripting languages to employ techniques such as parallel computing. Indeed, access to many big datasets (such as Twitter) are granted through APIs, and so they are primarily accessible to only those with programming skills. New techniques from computer science and statistics have allowed researchers to generate never-before seen insights (such as those from machine learning and artificial intelligence) from data that are otherwise very difficult to analyze. For instance, there are now a plethora of text-mining tools that were developed to harvest information from large quantities of textual records. By creating a pipeline that blends such methods with GIS it is possible to identify previously unexplored geospatial phenomena. Such techniques have also supported the growth of geographic data into wider disciplines.

There are dangers of simply recycling these methods for social research as they are inherently reductionist and lack sociological theoretical reasoning. In addition, we still need to undertake the most basic spatial operations, but with larger data. In response, we are now witnessing a re-emergence in the popularity of geocomputation due to the requirement to analyze large and complex datasets across space and time whilst retaining the basic principles of GIScience. There have been advancements on traditional techniques to enable their implementation on very large fractal datasets (Harris *et al.*, 2010). Open source examples of statistical programming languages which also extend to advanced spatial analysis include R and Python. In addition, there are Structured Query Languages (SQL) which are used to efficiently store and access large quantities of data in relational database management systems. In order to handle geospatial Big Data, many have utilized software such as PostGIS which specializes in integrating geographic objects into SQL databases. In many respects, there are parallels now to when GIS and computing first emerged and practitioners were required to code rather than click on buttons. It is important that such knowledge and experience are also integrated into academic programs to ensure that the skills of graduates are aligned with working in a data-driven world, in all its many forms.